\documentclass{article}
\usepackage[preprint, nonatbib]{neurips_2020}
\pdfoutput=1
\usepackage[utf8]{inputenc} % allow utf-8 input
\usepackage[T1]{fontenc}    % use 8-bit T1 fonts
\usepackage{hyperref}       % hyperlinks
\hypersetup{pdfauthor={Sharma, Gupta, Murty},pdftitle={Love tHy Neighbor}}
\usepackage{url}            % simple URL typesetting
\usepackage{booktabs}       % professional-quality tables
\usepackage{amsfonts}       % blackboard math symbols
\usepackage{nicefrac}       % compact symbols for 1/2, etc.
\usepackage{microtype}      % microtypography
\usepackage[numbers]{natbib}
\usepackage{hyperref}
\usepackage{subfig}
\usepackage{wrapfig}
\usepackage{graphicx}
\usepackage{booktabs}
\usepackage{xcolor}
\usepackage{amsmath}
\usepackage{amssymb}
\usepackage{amsthm}
\usepackage{wrapfig}
\usepackage{bm}

\usepackage[ruled, vlined, linesnumbered]{algorithm2e}
\usepackage{amsmath}
\usepackage{amssymb}
\usepackage{amsthm}
\usepackage{booktabs}
\usepackage{graphicx}
\usepackage{hyperref}
\usepackage[utf8]{inputenc}
\usepackage{soul}
\usepackage{subfig}
\usepackage{times}
\usepackage{url}
\usepackage{wrapfig}
\usepackage{xcolor}
\usepackage{dsfont}

\newtheorem{lemma}{Lemma}

\newcommand{\E}{E}
\newcommand{\G}{G}
\newcommand{\powerset}{\mathcal{P}}

\newcommand{\emailtag}{A}
\newcommand{\contacttag}{B}
\newcommand{\ndctag}{C}
\newcommand{\tagtag}{D}
\newcommand{\threadtag}{E}
\newcommand{\dblptag}{F}
\newcommand{\Hmax}{H\textsubscript{m}}
\newcommand{\Havg}{H\textsubscript{a}}
\newcommand{\HLone}{H\textsubscript{1}}
\newcommand{\HLtwo}{H\textsubscript{2}}

\title{Love tHy Neighbour: Remeasuring Local Structural Node Similarity in Hypergraph-Derived Networks}

\author{%
  Govind Sharma \qquad Paarth Gupta \qquad M. Narasimha Murty\\~\\
  \{\texttt{\href{mailto:govinds@iisc.ac.in}{govinds}, \href{mailto:paarthgupta@iisc.ac.in}{paarthgupta}, \href{mailto:mnm@iisc.ac.in}{mnm}\}@iisc.ac.in}\\~\\
  Department of Computer Science and Automation\\
  Indian Institute of Science, Bangalore\\
  Karnataka 560012, India
}

\begin{document}
\maketitle
\begin{abstract}
The problem of node-similarity in networks has motivated a plethora of such measures between node-pairs, which make use of the underlying graph structure. However, higher-order relations cannot be losslessly captured by mere graphs and hence, extensions thereof viz. hypergraphs are used instead. Measuring proximity between node pairs in such a setting calls for a revision in the topological measures of similarity, lest the hypergraph structure remains under-exploited. We, in this work, propose a multitude of hypergraph-oriented similarity scores between node-pairs, thereby providing novel solutions to the link prediction problem. As a part of our proposition, we provide theoretical formulations to extend graph-topology based scores to hypergraphs. We compare our scores with graph-based scores (over clique-expansions of hypergraphs into graphs) from the state-of-the-art. Using a combination of the existing graph-based and the proposed hypergraph-based similarity scores as features for a classifier predicts links much better than using the former solely. Experiments on several real-world datasets and both quantitative as well as qualitative analyses on the same exhibit the superiority of the proposed similarity scores over the existing ones.\end{abstract}

\section{Introduction}
  \label{sec:intro}
\begin{figure}[!t]
 \centering
 \begin{tabular}{ccc}
 \includegraphics[width=0.2\textwidth]{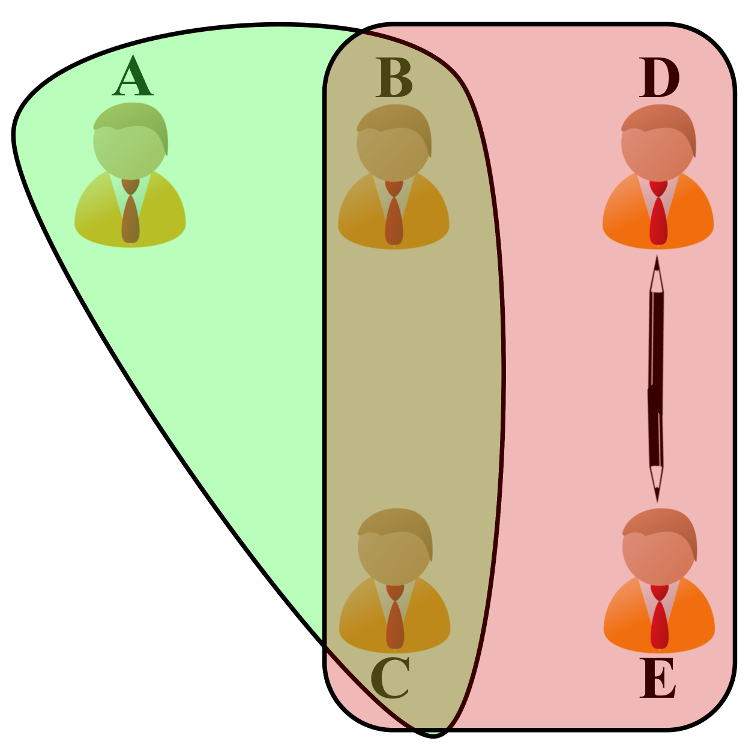} &
  \includegraphics[width=0.2\textwidth]{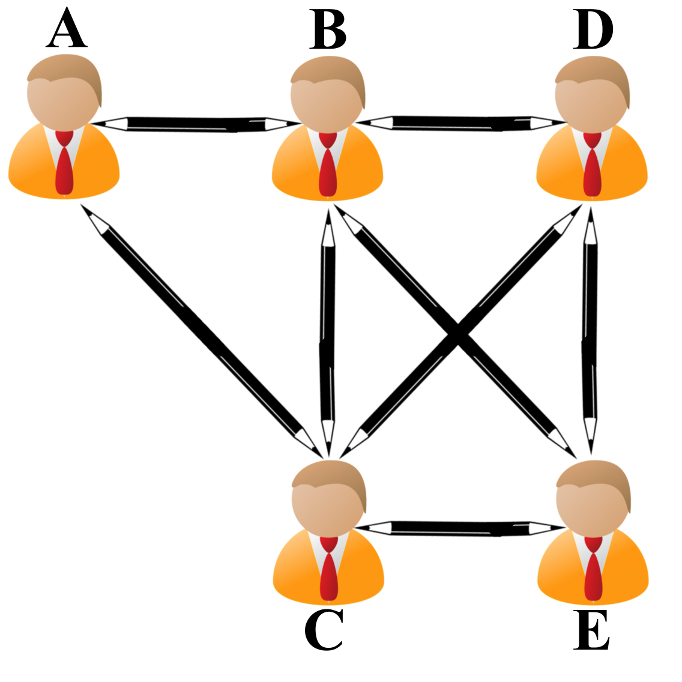} \\
  (a) Hypergraph &
  (b) Derived network
 \end{tabular}
 \caption{A toy example showing the genesis of a co-authorship network from its originally occurring hypergraph (\textit{i.e.}, its higher-order counterpart).}
 \label{fig:example}
%  \vspace{-1cm}
\end{figure}

  Measuring similarity between nodes of a graph has attracted the attention of network science researchers in all domains, be it social~\cite{gou2010social}, biological~\cite{bass2013using}, bibliographic~\cite{sun2011co}, or entertainment~\cite{kwon2012personalized}.
One simple reason why similarity between two nodes is important is to make a decision as to whether two seemingly unconnected nodes should be connected or not -- a problem more popularly known as recommendation~\cite{fouss2007random}.
While the notion of similarity between two nodes is fairly intuitive when the underlying relational structure of the network is graph-like (\textit{i.e.}, edges connect two nodes), it is a different ball game altogether when it is not.
More specifically, if the underlying relational structure of a network involves more than two entities in a single relation, the usual graph paradigm becomes lossy.
Moreover, it is quite unclear how close or similar two nodes would be in the presence of ``edges'' of higher sizes.
To make these two points clearer, let us divert our attention to Figure~\ref{fig:example}(a).
We see five authors $A$--$E$ who are related to each other by \textit{co-authorship}, which by nature possesses a higher-order property in that more than two authors can write a publication together.
In this case, we see three co-authorship groups: $ABC$, $BCDE$, and $DE$, each corresponding to a collaboration between the respective authors.
Our first point -- that we lose information when we pose these relations as pairwise -- is clear from Figure~\ref{fig:example}(b), which is a graph-induced version of the network.
Some illustrative pieces of information that the graph on the right loses are: (1) How many papers were written to start with? (2) Who all collaborated with each other? (3) Which author has the tendency to collaborate with larger teams?
Secondly, while it makes sense about how close or far away two nodes are to each other in a graph (say, authors $A$ and $E$ in Figure~\ref{fig:example}(b)), the same is not true for the relational structure on the left (Figure~\ref{fig:example}(b)).
The problem increases when we are asked to find the similarity between two nodes in such an scenario.
As would be shown later, standard measures of node similarity do not transfer directly from the graph-domain to hypergraphs~\cite{Berge1984,Bretto2013} -- so are structures capturing higher order relations called.

% Moreover, relational structure between pairs of entities is captured by a graph, which cannot handle higher-order relations.
\textit{Multiple authors authoring a paper together}, or \textit{multiple actors working on a movie project}, or \textit{multiple proteins interacting with each other}, \textit{etc.} are examples of such networks, and are better represented using \textit{hypergraphs} instead.
Hypergraphs are composed of arbitrary-sized edges called \textit{hyperedges}, which losslessly capture information from higher-order relations.
While most of the literature on node-similarity is focused on graphs, we deal with the same problem for hypergraphs\footnote{Please note that by ``node similarity on hypergraphs'', we still refer to pairs of nodes (\textit{links/edges}), \textbf{\textit{not}} hyperlinks/hyperedges.}.
Moreover, any given hypergraph could be converted into a graph (although lossily) by simple heuristics, \textit{weighted/unweighted clique-expansion}~\cite{Zhou2006,Agarwal2006} being one of them.
We call the resulting network a \textit{\textbf{hypergraph-expanded network}} or \textit{\textbf{hypergraph-derived network}}, that is, any network having an underlying hypergraph structure, \textit{i.e.}, higher-order relations between vertices.
The example in Figure~\ref{fig:example}(b) is such a network.
% \begin{definition}[Hypergraph Network (HNet)]
% \label{def:hnet}
%     Given a hypergraph $H = (V, F)$, where $F \subseteq \powerset(V)$ is the set of hyperedges over $V$, we define as an \textbf{HNet}, the network $G = G_H := (V, E)$, where $E = E_F := \left\{e \subseteq f~|~f \in F \text{ and } |e| = 2\right\}$ denotes the set of edges derived from $F$.
%     Optionally, we could define a weight $w(e) := |\{f \in F: e \subseteq f\}|$ on each edge.
% \end{definition}

% Some more examples of HNets have been listed in Table~\ref{tab:hnet_examples}.\\
% \textit{A note on non-HNets:} To understand HNets better, it is also useful to discuss an example of a ``non-HNet'' (or, say, a ``graph network'' or a ``GNet'').
% GNets could be defined as networks that do not have an underlying hypergraph structure and whose relations are originally pairwise.
% Some examples are friendship (person-person), road-/flight-connectivity (city-city), and commodity (locality-locality) networks, where the underlying relations are originally pairwise in nature, and not of a higher order.
% \begin{table*}[htb]
%  \caption{Examples of HNets $G_H = (V, E_F)$ derived from their underlying hypergraphs $H = (V, F)$}
%  \label{tab:hnet_examples}
%  \centering
%  \footnotesize
%  \input{tables/hypergraph_network_examples}
% \end{table*}
It is well-known that local similarity measures (\textit{e.g.}, \textit{common neighbors}~\cite{Newman2001}) prove to be powerful measures of node similarity, either by themselves~\cite{Liben-Nowell2003}, or as classifier features~\cite{AlHasan2006}.
However, while computing them, the underlying hypergraph structure remains hidden.
This deprives the algorithm of any extra information such higher-order structure could have contributed towards.
We \textbf{exploit the underlying hypergraph structure} and \textbf{extend popular local (i.e., neighborhood-based) similarity measures to their higher-order versions}.

In the present work, we focus on exploiting the topological properties of the underlying hypergraph of a derived network to aid the process of measuring the similarity between two nodes.
We restrict ourselves to local neighborhood based methods and argue using both information theory as well as via explicitly predicting links using the similarity scores.
Our experiments include performing experiments on both temporal as well as non-temporal datasets.

We first provide a generic formulation to convert any neighborhood based pairwise score to hypergraphs in Section~\ref{sec:formulation}.
Then in Section~\ref{sec:methodology}, we describe procedures to carefully prepare data and hypergraph-oriented features so as to carry out experiments.
In Section~\ref{sec:exp}, we perform experiments, that include both temporal as well as non-temporal datasets for the sake of completeness.
We also compute mutual information scores to vouch for the relevance of our measures, and devise different feature combinations to test them in a supervised learning scenario.
And finally, in Section~\ref{sec:results}, we compare our AUC scores with that of the baselines.
% \textbf{\textit{What we are not doing}}: \textit{We are \textit{\textbf{not}} dealing with the ``hyperlink prediction''~\cite{Xu2013} (or the ``hyperedge prediction'') problem in hypergraphs.}
% Lest the simultaneous mention of terms such as \textit{link}, \textit{edge}, \textit{pair}, \textit{node-similarity}, \textit{etc}.---even \textit{link prediction} for that matter---misleads the reader (esp. in the context of hypergraphs), we make attempts to clarify what we are not working on, in the present paper.  To elaborate further, we are interested in formulating the similarity (and thereby predicting the presence of a pairwise link) between \textit{\textbf{exactly two distinct nodes}} in a hypergraph network.
\subsection*{Our Contributions}
 \label{sec:intro:contri}
 \begin{enumerate}
  \item We formulate a \textbf{theoretically-backed novel technique} to convert graph topology-based pairwise node similarity measures into hypergraph-topology-based ones.
  \item We \textbf{extend} the local neighborhood based node similarity scores to their \textbf{hypergraph variants}.
  \item We propose \textbf{fair and unbiased novel data preparation} algorithms, so that similarity computation could be performed on both temporal and non-temporal hypergraph networks.
  \item We \textbf{improve the quality of structural similarity between nodes} by incorporating hypergraphs and the scores we formulate.
 \end{enumerate}

 \section{Background and Notation}
  \label{sec:bkgd}
  We define a \textit{hypergraph} by $H := (V, F)$, where $V$ denotes the set of its \textit{vertices}/\textit{nodes} and $F \subseteq 2^V$, its \textit{hyperedges}.
Its temporal variant, a \textit{temporal hypergraph} is one wherein for each hyperedge $f \in F$, information about the time of its (first) occurrence is communicated via a function $T_H: F \rightarrow \mathbb{R}$.
A \textit{graph}, on the other hand, is denoted by $G = (V, E)$ (with $(V, E, T_G)$ being its temporal variant), where $E \subseteq \powerset_2(V)$.
A hypergraph $H = (V, F)$ could be converted into a graph $\xi(H) := (V, E)$ with $E$ defined as $\cup_{f \in F} \powerset_2(f)$, a process known as \textit{clique expansion}~\cite{Agarwal2006}.
Refer to Table~\ref{tab:notation} for a quick reference to notations.
\begin{table}[htb]
    \caption{Notations used in this article}
    \label{tab:notation}
    \centering
    \begin{tabular}{rl}
       \toprule
        \textbf{Symbol} & \textbf{Definition} \\ 
        \toprule
        $V$ & Set of nodes \\
        $\powerset_k(\cdot)$ & $k$-power set of a set \\
        $F \subseteq 2^V$ & Set of hyperedges \\
        $E \subseteq \powerset_2(V)$ & Set of edges \\
        $H = (V, F)$ & Hypergraph \\
        $G = (V, E)$ & Graph \\
        $S \in \mathbb{R}^{|V| \times |F|}$ & Incidence Matrix \\
        $A \in \mathbb{R}^{|V| \times |V|}$ & Adjacency Matrix \\
        $\Gamma: V \rightarrow 2^V$ & Set of neighbors of a node\\
        $\tilde{\Gamma}: V \rightarrow 2^F$ & Set of hyperneighbors of a node\\
        % $\Phi \ni \varphi: \powerset_2(\powerset(V)) \rightarrow \mathbb{R}$ & Set-similarity function \\
        % $\Omega \ni \omega: \powerset_2(V) \rightarrow \mathbb{R}$ & Node-similarity function \\
        % $\mathcal{M}_\mathbb{R}$ & All real-valued finite matrices \\
        % $\Psi \ni \psi: \powerset_2(V) \rightarrow \mathcal{M}_\mathbb{R}$ & Node-similarity matrix-function \\
        % $\alpha: \Phi \rightarrow \Omega$ & Adjacency functional \\
        % $\eta: \Phi \rightarrow \Psi$ & Incidence matrix-functional \\
        % $\|\cdot\|: \mathcal{M}_\mathbb{R} \rightarrow \mathbb{R}$ & Matrix norm \\
        % $\|\eta\|: \Phi \rightarrow \Omega$ & Incidence functional \\
        \bottomrule
    \end{tabular}
\end{table}

% \footnote{The notion of ``expansion'' could be extended for a concept $X$ as: $\xi(X) = \powerset_2(X)$ if $X$ is a hyperedge, $\bigcup_{A \in X}\powerset_2(A)$ if $X$ is a set of hyperedges, and $(V, \xi(F))$ if $X$ is a hypergraph $(V, F)$.}

 \section{Formulating Similarities}
  \label{sec:formulation}
  Social scientists have long been involved in finding metrics that describe relations between entities in a network.
It was Liben-Nowell, et al.~\cite{Liben-Nowell2003} who first formally accumulated topological similarity scores from the network science literature and showed that they are good measures by themselves.
These similarity functions range from the earliest works of Katz~\cite{Katz1953} and Adamic, et al~\cite{Adamic2003} to the then recent works by Newman, et al.~\cite{Newman2004}; and till date, topological similarity computation features have witnessed several advancements~\cite{Martinez2016}.
Nevertheless, though there have been several works pertaining to hypergraphs and their applications, no work in the literature (with exceptions to some works~\cite{Li2013}, which use uniform/heterogeneous hypergraphs) utilizes usual (non-uniform, homogeneous) hypergraphs for similarity computation.

In this section, we set out to formally extend similarity computation scores in the literature from the graph- to the hypergraph-domain.
For the same, we define an end-to-end process of carefully constructing such scores from existing graph-topology based ones.
We first generalize graph-based (esp., local neighborhood based) scores via defining set-similarity functions taken from well-known local similarity computation paradigms, and then extend them to graphs (which is a usual, adjacency-based topology) and finally to a hypergraph (incidence-based) topology.

Our ultimate goal is to be able to predict links between an unlinked pair $\{u, v\}$ of nodes given a hypergraph $H$.
But in the literature, we find a plethora of techniques that make use of the existing graph structure, given graph $G$.
Most such techniques are set-based, in that they take sets $S_u, S_v$ corresponding to the nodes $u, v$ in consideration, and assign a prediction score to the pair $\{u, v\}$.

\subsection{The Case of Common-Neighbors}
For a pair of nodes $\{u, v\}$, the common-neighbors (CN) technique takes sets $S_u$ and $S_v$ to be \textit{neighbors} of $u$ and $v$ respectively, and computes the cardinality of their intersection.
In essence, CN makes use of two major concepts: ``neighborhood'', and ``intersection''.
Now, if one were to compute the hypergraph-equivalent to CN, one would have to use a concept equivalent to ``neighborhood''.
A simple option would be to consider ``hyperneighborhood'' instead (ref. Table~\ref{tab:notation}).
In other words, the hypergraph equivalent of CN for $u, v$ could be defined as the \textit{number of hyperedges incident on both $u$ and $v$}; but since the nodes are unlinked in the first place, there would be no common hyperedges!
Thus, this option \textit{fails}.
However, we still want to use the ``common-neighbor'' paradigm on hypergraphs.

Hence, first consider pairs of hyperedges $f_u, f_v \in F$, one incident to $u$ ($u \in f_u$) and the other to $v$ ($v \in f_v$), and count their ``intersection'' $|f_u \cap f_v|$.
And since $f_u$ and $f_v$ are precisely the hyperneighbors ($\tilde\Gamma$) of $u$ and $v$, we have combined the two concepts of ``hyperneighborhood'' and ``intersection'', and thus, have extended CN to hypergraphs.
But since each choice of $f_u, f_v$ would give a number $|f_u \cap f_v|$, we would have an $|\tilde\Gamma(u)| \times |\tilde\Gamma(v)|$ matrix of intersection-counts.
A suitable \textit{matrix norm} could then be used to convert this matrix into a single number, which would be used as a feature for similarity computation.

Extending this formalization to all local-neighborhood~\cite{Guns} based similarity computation scores is the ultimate goal of this section.
In order to do that, we define the notions of a \textit{set similarity function} $\varphi$, a \textit{node similarity function} $\omega$, and a \textit{node-similarity matrix-function} $\psi$.
For ease of transference of any set-similarity notion to node pairs, we also define two functionals (functions that map functions to functions): an \textit{adjacency functional} $\alpha$ (to use set-similarity functions as similarity computers in graphs) and an \textit{incidence functional} $\|\eta\|$ (to use them as similarity computers in hypergraphs).
An intermediate concept: \textit{incidence matrix-functional} $\eta$ has to be defined, so that it could be composed with a matrix norm to obtain incidence-based node-similarity measures.

\subsection{Extending Similarities to Hypergraphs}
Given a hypergraph $H = (V, F)$, for each vertex pair $\{u, v\} \in \powerset_2(V)$, we define functions  that quantify the proximity between vertices $u$ and $v$.

\begin{table}[!b]
    \caption{Set Similarity Functions for a graph $G = (V, E)$. Here, $U, U^\prime \subseteq V$ and $\Gamma(u)$ represents the neighbors of a node $u \in V$.}
    \label{tab:set_similarity_functions}
    \centering
    \begin{tabular}{lc}
       \toprule
        Common Neighbor, $\varphi_{CN}$ & $|U \cap U^\prime|$ \\
        \midrule
        Jaccard Coefficient, $\varphi_{JC}$ & $({|U \cap U^\prime|})/({|U \cup U^\prime|})$ \\
        \midrule
        Association Strength, $\varphi_{AS}$ & $({|U \cap U^\prime|})/({|U| \cdot |U^\prime|})$ \\
        \midrule
        Cosine Similarity, $\varphi_{CS}$ & $({|U \cap U^\prime|})/({\sqrt{|U| \cdot |U^\prime|}})$ \\
        \midrule
        NMeasure, $\varphi_{NM}$ & $({|U \cap U^\prime|})/({\sqrt{|U|^2 + |U^\prime|^2}})$ \\
        \midrule
        MinOverlap, $\varphi_{MnO}$ & $({|U \cap U^\prime|})/({\min\{|U|,  |U^\prime|\}})$ \\
        \midrule
        MaxOverlap, $\varphi_{MxO}$ & $({|U \cap U^\prime|})/({\max\{|U|,  |U^\prime|\}})$ \\
        \midrule
        Adamic Adar, $\varphi_{AA}$ & $\sum_{u \in U \cap U^\prime} {1}/{\log(|\Gamma(u)|)}$ \\
        \midrule
        Pearson Correlation, $\varphi_{PC}$ & $\frac{|V|\cdot|U \cap U^\prime| - |U|\cdot|U^\prime|}{\sqrt{\left(|V|\cdot|U| - |U|^2\right)\left(|V|\cdot|U^\prime| - |U^\prime|^2\right)}}$ \\
        \midrule
        Preferential Attachment, $\varphi_{PA}$ & $|U| \cdot |U^\prime|$ \\
        \bottomrule
    \end{tabular}
\end{table}

Define a \textbf{set similarity function} as $\varphi: \powerset_2(2^V) \rightarrow \mathbb{R}$ as a function that assigns to an unordered pair of vertex sets, $U, U^\prime \in 2^V$, a real number $\varphi(\{U, U^\prime\})$ corresponding to a measure of similarity between the sets.
A list of known set-similarity functions has been included in Table~\ref{tab:set_similarity_functions}.
Let $\Phi := \mathbb{R}^{\powerset_2(2^V)}$ represent all set-similarity functions over $V$.
We then define a \textbf{node similarity function} as $\omega: \powerset_2(V) \rightarrow \mathbb{R}$ that assigns to a pair of nodes $u, v \in V$, a similarity score $\omega(\{u, v\}) \in \mathbb{R}$.
Let $\Omega := \mathbb{R}^{\powerset_2(V)}$ denote the set of all node-similarity functions over $V$.
At this point, we also define an \textbf{adjacency functional} $\alpha: \Phi \rightarrow \Omega$ that maps each set-similarity function $\varphi \in \Phi$ to a node-similarity function $\omega = \alpha(\varphi) \in \Omega$ defined as $\alpha(\varphi)(\{u, v\}) := \varphi(\{\Gamma(u), \Gamma(v)\})$.

In order to extend this successfully to hypergraphs, we define a \textbf{node-similarity matrix-function} as $\psi: \powerset_2(V) \rightarrow \mathcal{M}_{\mathbb{R}}$ to be one that assigns to a pair of nodes $u, v \in V$, multiple similarity scores arranged in a real matrix $\psi(\{u, v\}) \in \mathcal{M}_{\mathbb{R}}$.
Let the set of all such functions for $V$ be denoted by $\Psi := \mathcal{M}_\mathbb{R}^{\powerset_2(V)}$, where $\mathcal{M}_\mathbb{R} := \bigcup_{m, n \in \mathbb{N}} \mathbb{R}^{m \times n}$ denotes the set of all real-valued finite-dimensional matrices.
Then for a hypergraph, we define an \textbf{incidence matrix-functional} $\eta: \Phi \rightarrow \Psi$ that maps each set-similarity function $\varphi \in \Phi$ to a node-similarity matrix-function $\psi = \eta(\varphi) \in \Psi$ defined as
 \[
  \eta(\varphi)(\{u, v\}) := 
  \begin{bmatrix}
  \varphi(f_1, f_1^\prime) &  \cdots & \varphi(f_1, f_n^\prime) \\
  \vdots & \ddots & \vdots \\
  \varphi(f_m, f_1^\prime) & \cdots & \varphi(f_m, f_n^\prime)
  \end{bmatrix} \in \mathbb{R}^{m \times n},
 \]
As discussed above, multiple matrix norms could be used to convert this matrix to a real number. Some of them are:
(i) \textit{Max-norm}: $\|X\|_{max} := \max_{i, j}\{X_{ij}\}$, (ii) \textit{Avg-norm}: $\|X\|_{avg} := \frac{1}{m\cdot n}\sum_{i, j} X_{ij}$, (iii) \textit{L1-norm}: $\|X\|_{1} := \sum_{i, j} |X_{ij}|$, and (iv) \textit{L2-norm}: $\|X\|_{2} := \sqrt{\sum_{i, j} |X_{ij}|^2}$.

Finally, the composition of a matrix norm with the incidence matrix-functional forms an \textbf{incidence functional}, defined as $\eta: \Phi \rightarrow \Psi$, which ultimately gives a functional $\|\eta\|: \Phi \rightarrow \Omega$, defined as $\varphi \mapsto \|\eta\|(\varphi) := \|\eta(\varphi)\| \in \Omega$, mapping pairs $\{u, v\} \in \powerset_2(V)$ to incidence-based similarities $\|\eta(\varphi)(\{u, v\})\| \in \mathbb{R}$.

Of the functionals defined above, the adjacency and the incidence functionals make use of \textit{neighbors} and \textit{hyperneighbors} respectively to transfer set-similarity functions to node-similarity in graphs and hypergraphs respectively.
\subsubsection{Illustration with Common Neighbors}
\label{sec:formulation:ins}
For the sake of further clarity, we demonstrate the case for Common Neighbors.
We first pick $\varphi_{CN} \in \Phi$ (as defined in Table~\ref{tab:set_similarity_functions}) as the set similarity function.
The adjacency functional $\alpha$ maps $\varphi_{CN}$ to a node similarity function $\omega_{CN} := \alpha(\varphi_{CN}): \powerset_2(V) \rightarrow \mathbb{R}$, defined by $\{u, v\} \mapsto \varphi_{CN}(\Gamma(u), \Gamma(v)) := |\Gamma(u) \cap \Gamma(v)|$, the usual common-neighbor criterion for similarity computation in graphs.
Moving to the incidence (hypergraph) domain, we first use the incidence matrix-functional $\eta$ to map $\varphi_{CN}$ to a node-similarity matrix-function $\psi_{CN} := \eta(\varphi_{CN}): \powerset_2(V) \rightarrow \mathcal{M}_\mathbb{R}$.
If $\tilde\Gamma(u) = \{f_1, \hdots, f_m\}$ and $\tilde\Gamma(v) = \{f^\prime_1, \hdots, f^\prime_n\}$, then $\eta(\varphi_{CN})(\{u, v\})$ is a matrix whose $(i,j)^{\text{th}}$ entry would be .
If a matrix norm such as $\|\cdot\|_{max}$ is chosen, we get the incidence functional, $\|\eta\|_{max}$, that gives us the node similarity function $\omega_{HCNM}(\{u, v\})$ as
\begin{equation*}
\|\eta\|_{max}(\varphi_{CN})(\{u, v\}) = \displaystyle \max_{(f, f^\prime) \in \tilde{\Gamma}(u) \times \tilde{\Gamma}(v)} \{|\varphi_{CN}(\{f, f^\prime\})|\},
\end{equation*}
where $\varphi_{CN}(f_i, f_j^\prime) = |f_i \cap f_j^\prime|$.

\subsection{Sanity Check for Hypergraph-based Similarities}
For the sake of establishing the sanity of the recently developed mechanism, we have the following lemma.
\begin{lemma}
The Common-Neighbor set similarity function $\varphi_{CN}$, when used to define an incidence-based node similarity function $\|\xi(\varphi)\|$, for a graph $\G = (V, \E)$, assigns to each pair $\{u, v\} \in \powerset_2(V)$, a similarity score that is proportional to a constant power of the original score.
That is,
\begin{equation}
    \|\xi(\varphi)\|(\{u, v\}) = \lambda \cdot \left(\alpha(\varphi)(\{u, v\})\right)^{\beta},
\end{equation}
for at least one matrix norm $\|\cdot\|$, and for some scalars $\lambda, \beta > 0$.
\end{lemma}
\begin{proof}
% Let us take $\varphi_{CN}$ (Common Neighbors) from Table~\ref{t:c:lp:set_similarity_functions} as an example.
Suppose we have $\G = (V, \E)$ as a usual undirected graph.
For nodes $u, v \in V$ s.t. $u \neq v$, if $\Gamma(u) := \{x_1, x_2, \hdots, x_m\}$ and $\Gamma(v) := \{y_1, y_2, \hdots, y_n\}$, we get hyperneighbors
\begin{align}
    \tilde\Gamma(u) =& \{\{u, x\} ~|~ x \in \Gamma(u)\} \\
    =& \{\{u, x_1\}, \{u, x_2\}, \hdots, \{u, x_m\}\}.
\end{align}
Similarly,
\begin{equation}
    \tilde\Gamma(v) = \{\{v, y_1\}, \{v, y_2\}, \hdots, \{v, y_n\}\}.
\end{equation}
Now, if we take the matrix norm to be L1 ($\|\cdot\|$), we have:
\begin{align}
\nonumber\left\|\xi(\varphi_{CN})(\{u, v\})\right\|_1 =& \left\|
\left(
\varphi_{CN}(\{\{u, x_i\}, \{v, y_j\}\})
\right)_{\substack{1\leq i\leq m\\1\leq j\leq n}}
\right\|_1\\
\nonumber=& \left\|\left(|
\{u, x_i\}\cap \{v, y_j\}|\right)_{\substack{1\leq i\leq m\\1\leq j\leq n}}
\right\|_1\\
\nonumber= & \sum_{\substack{1\leq i\leq m\\1\leq j\leq n}}\mathds{1}(x_i=y_j) = |\Gamma(u) \cap \Gamma(v)|\\
= & ~\varphi_{CN}(\{\Gamma(u), \Gamma(v)\}) = \alpha(\varphi_{CN})(\{u, v\})
\end{align}

Taking different matrix norms, we get scores as shown in the table below (Table~\ref{t:c:lp:sanity_norms}).
\begin{table}[!htb]
\caption{Similarity scores between $u$ and $v$ ($\varphi_{CN}$) when hypergraph is actually a graph.}
    \label{t:c:lp:sanity_norms}
    \centering
    \begin{tabular}{ll}
    \toprule
    \textit{Norm} & $\left\|\xi(\varphi_{CN})(\{u, v\})\right\|$ \\
    \toprule
    $\|\cdot\|_{max}$ & $\mathds{1}(\varphi_{CN}(\{\Gamma(u), \Gamma(v)\}) > 0)$\\
    $\|\cdot\|_{avg}$ &  $\displaystyle\frac{1} {|\Gamma(u)|\cdot |\Gamma(v)|} \cdot \varphi_{CN}(\{\Gamma(u), \Gamma(v)\})$\\
    $\|\cdot\|_{1}$ &    $\varphi_{CN}(\{\Gamma(u), \Gamma(v)\})$\\
    $\|\cdot\|_{2}$ & $\sqrt{\varphi_{CN}(\{\Gamma(u), \Gamma(v)\})}$\\
    \bottomrule
    \end{tabular}
\end{table}

It could be observed that when $\|\cdot\|_1$ is used as matrix norm, $\varphi_{CN}$ becomes the same for both hypergraphs and graphs (\textit{i.e.}, $\lambda = \beta = 1$).
Scores from the other norms act as extra features that we get as a result of the ``incidence matrix'' interpretation of a graph.
The same procedure when repeated for $\varphi_{AA}$, $\varphi_{JC}$, and other similarity scores gives us either the same graph score, or a scalar multiple of a power of it.
\end{proof}
\noindent \textbf{Note}: It needs to be understood that \textit{complete equality is not required}, since ultimately, we use graph features along with hypergraph ones (macro/micro combinations for GH, WH, etc.).
Also, the hypergraph based scores act as new features that come from the incidence matrix interpretation of the hypergraph (even if it is a graph).

% \subsection{Factorization based Node Embeddings}
% \subsubsection{Adjacency Matrix Factorization}
% \begin{equation}
%  A = WW^T,
% \end{equation}
% where $W \in \mathbb{R}^{n \times k}$, $k$ representing the dimensionality of the latent space.
% \begin{equation}
%  \sigma_{AMF}(u, v) = W_u^TW_v,
% \end{equation}
% where $W_u, W_v \in \mathbb{R}^k$ are transposed row vectors (\textit{i.e.}, column vectors) corresponding to vertices $u$ and $v$ respectively.
% \subsubsection{Incidence Matrix Factorization}
% \begin{equation}
%  S = XY^T,
% \end{equation}
% where $X \in \mathbb{R}^{n \times k}$ and $Y \in \mathbb{R}^{m \times k}$ represent latent ($k$-dimensional) representations for nodes $V$ and hyperedges $H$ respectively.
% \begin{equation}
%  \sigma_{IMF}(u, v) = X_u^TX_v,
% \end{equation}
% where $X_u, X_v \in \mathbb{R}^k$ represent vectorial representations of nodes $u$ and $v$ respectively.
% \subsection{Neural Node Embeddings}
% Similar to matrix factorizations of $A$ and $S$ in the previous section, we embed nodes in both the adjacency- and the incidence-domain using autoencoders, and use heuristics to combine them for each pair, that ultimately gives us a similarity score between the nodes in the pair.
% \subsubsection{Adjacency Based}
% We use adjacency based features per node for the autoencoder, that tries to embed nodes based on 
% \subsubsection{Incidence Based}
 
 \section{Methodology}
  \label{sec:methodology}
  \subsection{Data Preparation and Preprocessing}
\label{sec:methodology:data}
% \begin{figure*}[thb]
%     \centering
%     \input{images/data_preparation}
%     \caption{Data preparation pipeline for link prediction on graph $G$.}
%     \label{fig:data_preparation}
% \end{figure*}
% \begin{figure*}[thb]
%     \centering
%     \input{images/data_preparation_hyg}
%     \caption{Data preparation pipeline for link prediction on hypergraph $H$.}
%     \label{fig:data_preparation}
% \end{figure*}

% \begin{algorithm}[htb]
% \caption{\textsc{preparesimilarity computationData}($G$, $r$, $p$) for graph data}
%  \input{algorithms/data_prep}
% \end{algorithm}

%\begin{algorithm}[htb]
%\caption{\textsc{preparesimilarity computationData}($H$, $r$, $p$) for hypergraph data}
%\label{alg:prepare_lp_data}
 %\input{algorithms/data_prep_hyg}
%\end{algorithm}

% \begin{algorithm}[htb]
% \caption{\textsc{temporalSplit}($G$, $r$) for graph data}
%  \input{algorithms/split_train_test}
% \end{algorithm}

% \begin{algorithm}[htb]
% \caption{\textsc{temporalSplit}($H$, $r$) for hypergraph data}
% \label{alg:temporal_split}
%  \input{algorithms/split_train_test_hyg}
% \end{algorithm}

\begin{algorithm}[htb]
\caption{\textsc{structuralSplit}($H$, $r$) for hypergraph data}
\label{alg:structural_split}
 \SetCommentSty{mycommfont}
\SetKwInOut{Input}{Input}
\SetKwInOut{Output}{Output}
\Input{
    Hypergraph $H = (V, F)$\newline
    Split ratio $r \in [0, 1]$
}
\Output{
    Train hyperedges $F_{tr}$\newline
    Test links $E_{te}$
}
\DontPrintSemicolon
$E \leftarrow \{\}$\; %\tcp*{edges $E$ initialized to be empty-set}
\For{$f \in F$}{
    \For{$e \in \powerset_2(f)$}{
        \If{$e \notin E$}{
            $E \leftarrow E \cup \{e\}$\;
        }
    }
}
$E_{te} \leftarrow$ \textsc{sample}($E$, $\lceil(1-r)\cdot |E|\rceil$)\;
$F_{tr} \leftarrow$ \textsc{cleanHyperedges}($F$, $E_{te})$\;
% $F^{(2)}_{tr} \leftarrow \{ \}$\;
% \For{$f \in F$}{
%     $E_f \leftarrow \powerset_2(f)$\;
%     $E_{te\wedge f} \leftarrow E_{te} \cap E_f$\;
%     \uIf{$E_{te\wedge f} = \emptyset$}{
%         $F^{(1)}_{tr} \leftarrow F^{(1)}_{tr} \cup \{f\}$\;
%         $F^{(2)}_{tr} \leftarrow F^{(2)}_{tr} \cup \{f\}$\;
%     }
%     \Else{
%         $F_{f-te} \leftarrow$ \textsc{cleanHyperedges}($\{f\}$, $E_{te\wedge f}$)\;
%         $F^{(2)}_{tr} \leftarrow F^{(2)}_{tr} \cup F_{f-te}$\;
%     }
% }
\Return{$F_{tr}$, $E_{te}$}
\end{algorithm}

\begin{algorithm}[htb]
\caption{\textsc{cleanHyperedges}($F$, $E$) to remove edge-information in $E$ from hyperedges $F$}
\label{alg:clean_hyperedges}
 \SetCommentSty{mycommfont}
\SetKwInOut{Input}{Input}
\SetKwInOut{Output}{Output}
\Input{
    Set of hyperedges $F$\newline
    Set of edges $E$
}
\Output{
    Cleaned-up hyperedges $F_{-E}$
}
\DontPrintSemicolon
$EF: E \rightarrow 2^F$\;
$F_{-E} \leftarrow F$\;

\For{$\{u, v\} \in E$}{
    $EF[\{u, v\}] \leftarrow \{f \in F ~|~ u, v \in f\}$\;
}
\For{$\{u, v\} \in E$}{
    \For{$f \in EF[\{u, v\}]$}{
        $f_{-u} \leftarrow f\setminus\{u\}$\;
        $f_{-v} \leftarrow f\setminus\{v\}$\;
        $F_{-E} \leftarrow (F_{-E}\left. \setminus \{f\})\right. \cup \{f_{-u},f_{-v} \}$\;
        $EF[\{u, v\}] \leftarrow EF[\{u, v\}] \setminus \{f\}$\;
        \For{$w \in f\setminus\{u, v\}$}{
            \If{$\{u, w\} \in E$}{
                $EF[\{u, w\}] \leftarrow \left(EF[\{u, w\}] \setminus \{f\}\right) \cup \{f_{-v}\}$\;
            }
            \If{$\{v, w\} \in E$}{
                $EF[\{v, w\}] \leftarrow \left(EF[\{v, w\}] \setminus \{f\}\right) \cup \{f_{-u}\}$\;
            }
        }
    }
}
\Return{$F_{-E}$}
\end{algorithm}

Given a hypergraph (temporal or non-temporal), we need to convert it into a form that is consumable in the similarity computation setting, so that we are able to calculate both graph- and hypergraph-based features readily.
We prepare data separately for temporal and structural similarity computation settings.
\subsubsection{Temporal Processing}
\label{sec:methodology:data:temporal}
In the temporal setting, we have a timed hypergraph, $H = (V, F, T_H)$ with us.
Unweighted and weighted clique expansions of $H$ give $G = (V, E, T_G)$ and $G_w = (V, E, T_G, w)$ respectively.

In short, we have timed graphs $G$ and $G_w$ with us now.
Now, a \textit{split-ratio} parameter $\rho \in [0, 1]$ is selected. The graph timeline (image of $T$ under $E$) $T(E) := \{t_1, \hdots, t_{n_T}\} \subseteq \mathbb{R}$ could be defined as a set of $n_T$ time-stamps for which, \textit{w.l.o.g.}, let $t_1 < \cdots < t_{n_T}$.
If a time-threshold index is now defined as $\tau := \lceil (1 - \rho) \cdot n_T \rceil$, we could divide the timeline into train- and test-periods\footnote{Please note that the terms `train period' and `test period' is akin to the formulation by Liben-Nowell, et al.~\cite{Liben-Nowell2003} and should not be confused with train and test datasets in supervised learning.} $T_{tr} := \{t_1, \hdots, t_\tau\}$ and $T_{te} := \{t_{\tau+1}, \hdots t_{n_T}\}$ respectively.
This leads us to edge sets $E_{tr} := \{e \in E ~|~ T(e) \in T_{tr}\}$ and $E_{te} := \{e \in E ~|~ T(e) \in T_{te}\}$, denoting edges formed in the train and test periods respectively.
Similarly, we define $F_{tr} := \{f \in F ~|~ T(f) \in T_{tr}\}$ as the hyperedges that were formed in the train period.

\subsubsection{Structural Processing}
\label{sec:methodology:data:temporal}
Structural processing is easier than temporal, since hyperedges are non-timed.
Formally, we start with a hypergraph $H = (V, F)$, which gets converted into graphs $G = (V, E)$ and $G_w = (V, E, w)$ as before.
A similar split-ratio $\rho \in [0, 1]$ is selected, and if $m := |E|$, we randomly delete $m_{te} := \lceil \rho \cdot m \rceil$ number of edges from the graph, which has to be predicted later.
In other words, a random sample $E_{te} \subseteq E$ is selected such that $|E_{te}| = m_{te}$.
As a result, we get the set of test edges $E_{te}$.

We now discuss the preparation of the train hypergraph, whose topology would be used while predicting links.
In the temporal case, we simply ignored hyperedges $F_{te}$ from the test period and what remained was $F_{tr}$.
But here, we have no temporal information, and the train-test split is done at random, which successfully separates $E_{tr}$ from $E_{te}$, but not $F_{tr}$ from $F_{te}$, since there are no well-defined concepts of ``train period'' or ``test period'' here.

Let us analyze the situation closely.
Before continuing further, let us extend the hyperneighborhood function to edges: $\tilde{\Gamma}: E \rightarrow 2^F$ defined as $e \mapsto\tilde\Gamma(e) := \{f \in F ~|~ e \subseteq F\}$.
The question is: which hyperedges should be included in the train set so that information from them could be used while predicting test links $E_{te}$?

% \textbf{Option 1}: \textit{Use all hyperedges}, i.e., $F^{(1)}_{tr} := F$.
% \begin{claim}
% \label{claim:1}
%  $\forall e \in E_{te}$, $\exists f \in F^{(1)}_{tr}$ such that $e \subseteq f$.
% \end{claim}
% That is, all edges that we set out to predict (those in $E_{te}$) would already be present in the form of their underlying hyperedges in the train hypergraph.
Choosing all hyperedges $F$ as $F_{tr}$ would \textit{\textbf{trivialize}} the very task of similarity computation and we would end up predicting all links with a 100\% accuracy using only one feature: ``common hyperneighbors''!
And, on the other hand, using only those hyperedges that are not supersets of any test edge, i.e., $F_{tr} = \{f \in F ~|~ f \not \supseteq e~ \forall e \in E_{te}\}$ would \textit{\textbf{deprive}} us of many links that a ``hyperedge \textit{minus} a test edge'' would have otherwise provided.
We go with neither of the options and choose to ``strip'' each test edge off of a potential train hyperedge.
A detailed procedure has been described in Algorithm~\ref{alg:structural_split}, which in turn uses Algorithm~\ref{alg:clean_hyperedges} to clean away information about any test edge from the hyperedges, finally giving us a rich train hypergraph for similarity computation.

Finally, we get train hypergraph $H_{tr} := (V, F_{tr})$, and test edges $E_{te}$.
The similarity computation problem would be to predict new links (i.e., those not already present in $E_{tr}$) using information from the hypergraph topology $H_{tr}$; predictions will later be evaluated using test set $E_{te}$.

% \textbf{Option 2}: \textit{Use only those hyperedges that are not supersets of any test edge}, i.e., $F^{(2)}_{tr} := \{f \in F ~|~ f \not \supseteq e~ \forall e \in E_{te}\}$.

% \begin{claim}
% \label{claim:1}
%  $\forall e \in E_{te}$, $\forall f \in F^{(2)}_{tr}$, we have $e \not\subseteq f$.
% \end{claim}

% Claim~\ref{claim:2} is precisely the negation of Claim~\ref{claim:1}.
% That is, no test edge would be present (either directly or in the form of an underlying hyperedge) in the train hypergraph.

% Let us now define $G_{tr}$, i.e., the train graph whose features we would have used to perform link prediction, had we considered only graph topology: $G_{tr} := (V, E_{tr})$, where $E_{tr} := E \setminus E_{te}$.

% $F_{tr}^{(2)}$, if we 

\subsection{Computing Graph Features}
\label{sec:methodology:graph_feats}
We had earlier listed certain set-similarity functions in Table~\ref{tab:set_similarity_functions}.
Let use take the corresponding link predictors (let us call them \textit{\textbf{base predictors}}) from the literature~\cite{Liben-Nowell2003,Guns} and hence get ten different similarity computation scores for each pair of nodes in a given dataset.
More specifically, we take the adjacency node similarity function $\alpha(\varphi_{i})$ where $\alpha$ and $\varphi$ are used as per Section~\ref{sec:formulation} (where base predictor, $i \in \{$\textit{AA}, \textit{JC}, \textit{AS}, \textit{CS}, \textit{NM}, \textit{MnO}, \textit{MxO}, \textit{AA}, \textit{PC}, \textit{PA}$\}$) to find scores for each pair.
We repeat this exercise for the edge-weighted version of the graph (using weighted scoring functions defined in~\cite{Guns}).
Finally, for each hypergraph, corresponding to each base predictor, we have two different graph-based topological scores per node pair, which we denote by \texttt{G} (for unweighted graph) and \texttt{W} (for weighted graph) respectively.

\subsection{Computing Hypergraph Features}
\label{sec:methodology:hypergraph_feats}
Similar to Section~\ref{sec:methodology:graph_feats}, also compute scores for the hypergraph-variations of the base-predictors.
This involves computing the node-similarity matrix-function $\psi$ for each of the set similarity functions mentioned in Table~\ref{tab:set_similarity_functions}, followed by the application of the four matrix norms defined earlier to obtain a single numeric score for each pair.
In summary, we compute the incidence node similarity function via $\|\eta(\varphi_{i})\|$, where $\varphi$, $\eta$, and $\|\|$ are as defined earlier, and Table~\ref{tab:set_similarity_functions}.
For each hypergraph, corresponding to each base predictor, we have four different hypergraph-based topological scores per node pair, which we denote by \texttt{\Hmax}, \texttt{\Havg}, \texttt{\HLone}, and \texttt{\HLtwo}.
 
%  \section{A Note on Data Preparation}
%   \label{sec:data_prep}
%   \input{sections/data_prep}
 \section{Related work}
  \label{sec:sota}
  Computing similarity scores has a vast literature, and covering it in whole is beyond the scope of the present work.
The reader is redirected to some excellent review works~\cite{Wang2015,Lu2010,Martinez2016}, which provide an intelligible coverage of the similarity computation ecosystem.
Although the concept wasn't new to network scientists, and there have been vintage works on predicting new relations in networks (\cite{Katz1953}), the first formal work on similarity computation could be credited to Liben-Nowell, et al.~\cite{Liben-Nowell2003}. 
They brought together multiple similarity scores to solve the problem, scores both new and existing~\cite{Katz1953,Adamic2003,Newman2001}.
Even since, many interesting directions to solve the similarity computation problem in networks were taken.

However, almost all works that use hypergraph networks (with the exception of Li et al.~\cite{Li2013}, who deal with heterogeneous, uniform hypergraphs only) do not consider the underlying hypergraph structure after the network gets expanded to a graph.
Recently, there has been interest in the areas of hyperlink (or simplex, or merely hyperedge) prediction as well, which acknowledges the fact that there is a loss of information when a hypergraph is converted to a graph~\cite{Xu2013,Benson2018}.
 
 \section{Experiments}
  \label{sec:exp}
  \subsection{Datasets}
\label{sec:exp:datasets}
We use a multitude of hypergraph datasets, mainly from Benson, et al.~\cite{Benson2018}, from where we pick six datasets.
A brief account of all of them is as follows:
\begin{itemize}
    \item [\emailtag] \textbf{email-Enron}: In an organization (Enron Corporation), an \textit{email communication} between \textit{employee} nodes represents a hyperedge~\cite{klimt2004introducing}.
    \item [\contacttag] \textbf{contact-high-school}: In a high-school setting, nodes represent \textit{school students}, and a hyperedge is formed between \textit{individuals that are spatially close} to each other at a given time instance~\cite{mastrandrea2015contact}.
    \item [\ndctag] \textbf{NDC-substances}: Nodes signify \textit{chemical substances}, and a hyperedge represents a set of these substances used in a particular \textit{drug}.
    \item [\tagtag] \textbf{tags-math-sx}: Again, it is a dataset from the same mathematical forum as above, only that the nodes denote \textit{mathematical tags}, and a hyperedge is formed over all \textit{tags that a particular question is associated with}.
    \item [\threadtag] \textbf{threads math-sx}: Users on a mathematics discussion forum\footnote{\textit{https://math.stackexchange.com/}} form nodes and a \textit{group of users involved in a particular question thread} forms a hyperedge.
    \item [\dblptag] \textbf{coauth-DBLP}: Nodes represent \textit{authors} and a hyperedge, a group of all \textit{authors that wrote a paper together}.
\end{itemize}
Refer to \cite{Benson2018} for more details.
\subsection{Preprocessing data and computing scores}
\label{sec:exp:preprocess}
We perform a lot of link-prediction experiments on a number of hypergraph datasets belonging to multiple real-world domains.
Since data preparation is both a crucial step as well as one of our main contributions, it forms a major part in our methodology (Section~\ref{sec:methodology:data}) itself.
We fix the split-ratio to be $r = 0.2$, and choose to \textit{randomly} generate $p = 5$ times as many negative samples (non-links) as positive samples (links).
For each hypergraph, we perform \textit{both} temporal and structual link prediciton (ignoring the time information for the latter).
We get train hyperedges $F_{tr}$, test links $E_{te}$, and test non-links $\hat{E}_{te}$ as defined above.

For each pair $\{u, v\} \in E_{te} \cup \hat E_{te}$, we compute the ten base predictor scores, as mentioned in Section~\ref{sec:methodology:graph_feats}, taking $E_{tr} := \xi(F_{tr})$ (both weighted and unweighted) as information for edges, hence preparing our baselines.
Then, as explained in Section~\ref{sec:methodology:hypergraph_feats}, we compute hypergraph-topology based scores that we have proposed.
Towards the end, for each base predictor, we have a total of six different scores per node pair: \textit{graph} (\texttt{G}), \textit{weighted-graph} (\texttt{W}), \textit{hypergraph-max} (\texttt{\Hmax}), \textit{hypergraph-avg} (\texttt{\Havg}), \textit{hypergraph-L1} (\texttt{\HLone}), and \textit{hypergraph-L2} (\texttt{\HLtwo}).
And since there are a total of ten base predictors: \textit{AA}, \textit{AS}, \textit{CN}, \textit{Cos}, \textit{PA}, \textit{JC}, \textit{MxO}, \textit{MnO}, \textit{NM}, and \textit{Prn}, we finally get $6 \times 10 = 60$ different scores per node pair.

\subsection{Calculating Mutual Information}
\label{sec:exp:mi}
Mutual information~\cite{shannon1948mathematical} has been shown to play a major role in similarity computation~\cite{tan2014link}.
But we use it here in the classical sense, in that for each dataset, we find the mutual information score for each individual feature by binning its values via a log-binning (where consecutive bins are assigned on the base-10 log scale) mechanism since they are continuous values, with all of them being power-law distributed as opposed to normal.
We monitored the MI scores for various number of bins and found that beyond a sufficiently large number of bins, the relative rank of the similarity computation features does not change.
Hence, we fix the number of bins to be 2000.

\subsection{Performing Link Prediction}
\label{sec:exp:lp}
Finally, we perform similarity computation in three different modes, which have been described as follows:
\begin{enumerate}
    \item \textit{\textbf{Standalone features}}: In this mode, we simply use the predictor scores (\texttt{G}, \texttt{W}, \texttt{\Hmax}, \texttt{\Havg}, \texttt{\HLone}, \texttt{\HLtwo}) calculated in Section~\ref{sec:exp:preprocess} for similarity computation, \textit{i.e.}, predict links via the unsupervised similarity computation paradigm similar to Liben-Nowell et al.~\cite{Liben-Nowell2003}.
    At the end, we would have a total of 60 standalone scores.
    Although we did not expect to do better than the baselines in this mode, we still observe decent performances).
% Going by a base predictor individually (row-wise), graph-versions (\texttt{std-G}) of Adamic Adar (AA) for structural- and Cosine Similarity (Cos) for the temporal-mode perform best.
    \item \textit{\textbf{Micro-feature combination}}: Here, we take various feature combinations, treating each of the ten base predictors separately.
    We have a total of five different feature combinations per base predictor: \texttt{mic-G}, \texttt{mic-W}, \texttt{mic-H}, \texttt{mic-GH}, \texttt{mic-WH}, where the first two correspond to singleton features \texttt{G} and \texttt{W}, and the last three to taking \texttt{H} individually, \texttt{G} and \texttt{H} together, and \texttt{W} and \texttt{H} together respectively (ref. Sections~\ref{sec:methodology:graph_feats} and~\ref{sec:methodology:hypergraph_feats}).
    In all, we have 10 $\times$ 5 = 50 micro feature combinations for each dataset.
    \item \textit{\textbf{Macro-feature combination}}: This is similar to micro-feature combination, except \textit{all} base predictors are taken together for each combination.
    That is, we take all graph-based features (\texttt{mac-G}), all weighted-graph-based features (\texttt{mac-W}), all hypergraph-based features (\texttt{mac-H}), and their combinations \texttt{mac-GH} and \texttt{mac-WH}.
    We have totally 5 macro-feature combinations for each dataset.
\end{enumerate}

In case of \textit{micro} and \textit{macro} modes, we learn an XGBoost~\cite{chen2016xgboost} classifier to predict links (and get one classifier per feature combination), and in the \textit{standalone} mode, the scores themselves are used as predictions.
For the classification, we randomly split the prepared data $E_{te} \cup \hat E_{te}$ further into train and test, this time for classification\footnote{Earlier, we had performed a train-test split in a temporal or a structural sense, which was a data preparation step. But here, the usual, supervised-learning oriented split of the \textit{prepared} data into train and test has been performed}.
Once we have the predictions by a feature combination, for evaluating performance, the predictions are compared with the labels (link/non-link) and ROC curves~\cite{Davis2006} are derived, which are finally summarized using Area Under ROC (AUC).

 \section{Results and Discussion}
  \label{sec:results}
  % \subsection{Node Pair Similarities}
% \label{sec:results:node_similarities}
% \input{sections/results/node_similarities.tex}

% \subsection{Link Prediction}
% \label{sec:results:lp}

% We perform\footnote{Note: We make the code to generate the results in this paper available online at \texttt{\url{https://tinyurl.com/u5kj7x5}}.} the experiments listed in the previous section on all the six datasets, all base predictors.
We perform the experiments listed in the previous section on all the six datasets, all base predictors.
For \textit{micro} and \textit{macro} modes, we get a total of 50 and 5 classifiers respectively (one per feature combination), and the same number of AUC scores, and for the \textit{standalone} mode, we have 60 different AUC scores.
Since owing to space limitations, it is difficult to show all the results here, we try our best to summarize all our experiments as best as possible using a handful of results.
We run these experiments for a total of \textit{five} times, so as to monitor the variance across different runs, since each experiment has at least one random step, \textit{viz.}, sampling of non-links.

\begin{figure*}%[thb]
    \centering
    \includegraphics[width=0.9\linewidth]{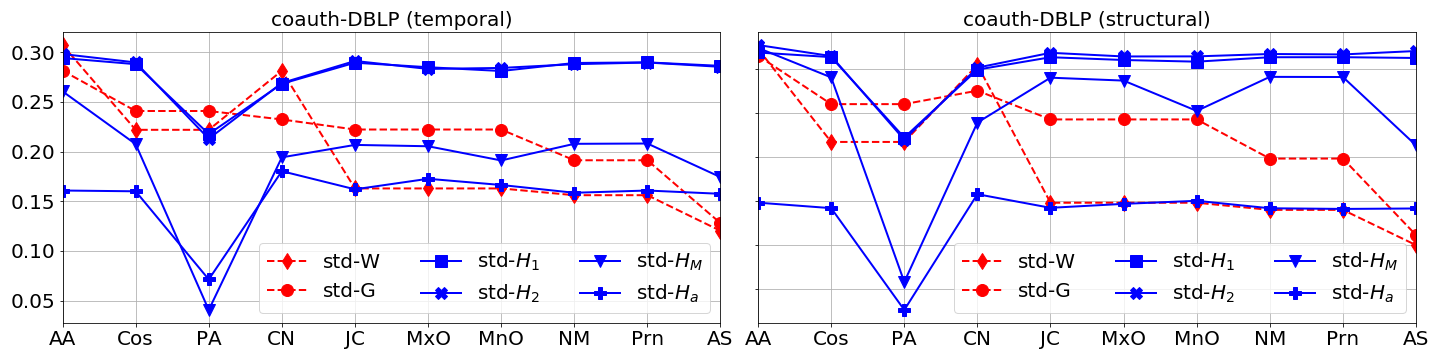}
    \caption{Mutual information scores denoting importance of six features (for all ten base predictors) in classifying links vs. non-links, computed on the coauth-DBLP hypergraph (dataset \dblptag)}
    \label{fig:DataSet:Coauth-DBLP}
\end{figure*}

\subsection{Mutual Information for Link Prediction}
Treating each standalone score as a feature in a supervised setting, we compute their mutual information (MI) \textit{w.r.t.} the positive (links) and negative (non-links) classes.
For the dataset coauth-DBLP, we plot MI scores for both temporal and structural similarity computation for each base predictor.
As could be observed, in the temporal case, except for AA, PA, and CN, where graph or weighted-graph MI outperforms the others, at least two hypergraph MI scores are better than the graph ones.
This only means that hypergraph based scores have the potential to better \textit{explain} links vs. non-links.
We chose this dataset since it is the largest hypergraph we have used.

\subsection{Micro Feature Combination Performances}
As per the description of the \textit{micro feature combination} mode in Section~\ref{sec:exp:lp}, we report AUC scores for the contact-high-school data in Table~\ref{tab:micro_feat_perf_auc_contact-high-school}.
It has to be interpreted as per various micro-feature combinations.
As is clear from the highlighted numbers, except for Cos, JC, and MxO in the temporal similarity computation case (which perform best with \texttt{mic-W}), in all other cases, feature combinations involving hypergraphs (\texttt{mic-H}, \texttt{mic-GH}, \texttt{mic-WH}) work best.

\begin{table}
    \caption{AUC scores (\%) for structural (-s) and temporal (-t) link prediction using micro-feature-combination via XGBoost for \textit{contact-high-school} (\textit{i.e.}, dataset \contacttag). Row ids AA--Prn represent base predictors.}
    \label{tab:micro_feat_perf_auc_contact-high-school}
    \centering
    \footnotesize
    \begin{tabular}{l@{\hspace{2mm}}c@{\hspace{2mm}}c@{\hspace{2mm}}c@{\hspace{2mm}}c@{\hspace{2mm}}c}
% \begin{tabular}{lccccc}
\toprule
{} & \texttt{mic-G} &         \texttt{mic-W} &         \texttt{mic-H} &        \texttt{mic-GH} &        \texttt{mic-WH} \\
\midrule
AA-s  &   93.0$\pm$0.7 &           92.8$\pm$0.8 &           93.3$\pm$0.6 &           \textbf{93.4$\pm$0.5} &  \textbf{93.4$\pm$0.6} \\
AS-s  &   91.5$\pm$0.8 &           88.3$\pm$0.6 &           93.3$\pm$0.4 &  \textbf{93.5$\pm$0.4} &           93.4$\pm$0.5 \\
CN-s  &   93.0$\pm$0.7 &           92.6$\pm$0.9 &           92.9$\pm$0.3 &  \textbf{93.2$\pm$0.4} &  \textbf{93.2$\pm$0.4} \\
Cos-s &   92.9$\pm$0.8 &           93.0$\pm$0.6 &           93.1$\pm$0.3 &           93.2$\pm$0.4 &  \textbf{93.5$\pm$0.5} \\
PA-s  &   62.3$\pm$0.9 &           60.9$\pm$1.6 &           62.3$\pm$1.5 &           62.6$\pm$1.2 &  \textbf{63.7$\pm$1.5} \\
JC-s  &   92.8$\pm$0.5 &           92.8$\pm$0.4 &           93.1$\pm$0.3 &           \textbf{93.3$\pm$0.2} &  \textbf{93.3$\pm$0.3} \\
MxO-s &   92.6$\pm$0.4 &           92.5$\pm$0.4 &           93.2$\pm$0.4 &  \textbf{93.3$\pm$0.4} &           \textbf{93.3$\pm$0.3} \\
MnO-s &   92.6$\pm$0.9 &           91.5$\pm$0.6 &           93.0$\pm$0.2 &  \textbf{93.3$\pm$0.7} &           93.1$\pm$0.3 \\
NM-s  &   92.8$\pm$0.5 &           92.5$\pm$0.3 &           93.2$\pm$0.3 &           93.3$\pm$0.3 &  \textbf{93.4$\pm$0.4} \\
Prn-s &   90.9$\pm$0.6 &           90.8$\pm$0.8 &           93.2$\pm$0.3 &           93.2$\pm$0.3 &  \textbf{93.3$\pm$0.4} \\
\midrule
AA-t  &   86.3$\pm$2.4 &           86.7$\pm$2.4 &           87.3$\pm$2.0 &           87.4$\pm$1.8 &  \textbf{87.9$\pm$2.2} \\
AS-t  &   85.9$\pm$1.3 &           84.0$\pm$1.4 &  \textbf{87.5$\pm$1.9} &           86.9$\pm$1.8 &           87.2$\pm$1.9 \\
CN-t  &   87.3$\pm$2.0 &           86.8$\pm$1.8 &           86.4$\pm$1.9 &           86.8$\pm$2.1 &  \textbf{87.4$\pm$2.1} \\
Cos-t &   87.5$\pm$1.6 &  \textbf{88.1$\pm$2.0} &           87.4$\pm$1.8 &           87.3$\pm$1.9 &           87.5$\pm$2.1 \\
PA-t  &   53.6$\pm$2.1 &           54.0$\pm$2.3 &           52.4$\pm$3.6 &           55.0$\pm$2.7 &  \textbf{57.2$\pm$3.2} \\
JC-t  &   87.5$\pm$1.9 &  \textbf{88.4$\pm$1.9} &           87.5$\pm$1.8 &           87.4$\pm$1.7 &           88.0$\pm$1.8 \\
MxO-t &   86.9$\pm$1.9 &  \textbf{87.7$\pm$1.3} &           87.4$\pm$1.5 &           87.2$\pm$1.6 &           87.5$\pm$1.3 \\
MnO-t &   86.9$\pm$1.2 &           86.6$\pm$1.5 &           86.6$\pm$2.2 &           86.5$\pm$1.9 &  \textbf{87.7$\pm$1.8} \\
NM-t  &   86.6$\pm$2.0 &           87.6$\pm$1.4 &           87.3$\pm$1.8 &           87.2$\pm$1.7 &  \textbf{87.9$\pm$1.8} \\
Prn-t &   84.0$\pm$2.6 &           83.7$\pm$2.2 &           87.3$\pm$2.0 &           87.1$\pm$2.2 &  \textbf{87.5$\pm$2.1} \\
\bottomrule
\end{tabular}
\end{table}
\begin{table}
    \caption{Rank-performances \textit{w.r.t.} AUC scores from Table~\ref{tab:micro_feat_perf_auc_contact-high-school} across all datasets.}
    \label{tab:micro_feat_rank_perf_auc}
    \centering
    \footnotesize
    \begin{tabular}{l@{\hspace{2.5mm}}c@{\hspace{2.5mm}}c@{\hspace{2.5mm}}c@{\hspace{2.5mm}}c@{\hspace{2.5mm}}c@{\hspace{2.5mm}}c}
\toprule
{} & \texttt{mic-G} & \texttt{mic-W} &        \texttt{mic-H} &       \texttt{mic-GH} &       \texttt{mic-WH} \\
\midrule
\emailtag-s    &    3.6$\pm$0.5 &    5.0$\pm$0.0 &           3.2$\pm$0.7 &  \textbf{1.4$\pm$0.5} &           1.8$\pm$0.7 \\
\contacttag-s &    4.0$\pm$0.4 &    4.9$\pm$0.3 &           3.0$\pm$0.4 &           1.7$\pm$0.5 &  \textbf{1.4$\pm$0.4} \\
\ndctag-s     &    4.0$\pm$0.2 &    5.0$\pm$0.2 &           2.3$\pm$0.5 &  \textbf{1.8$\pm$0.2} &  \textbf{1.8$\pm$0.2} \\
\tagtag-s     &    4.3$\pm$0.5 &    4.7$\pm$0.5 &           2.6$\pm$0.4 &  \textbf{1.3$\pm$0.5} &           2.1$\pm$0.4 \\
\threadtag-s  &    4.2$\pm$0.2 &    4.8$\pm$0.2 &  \textbf{1.9$\pm$0.3} &           2.0$\pm$0.2 &           2.0$\pm$0.2 \\
\dblptag-s    &    3.1$\pm$0.3 &    3.2$\pm$0.6 &           3.0$\pm$0.0 &  \textbf{2.8$\pm$0.4} &  \textbf{2.8$\pm$0.4} \\
\midrule
\emailtag-t    &    4.8$\pm$0.4 &    4.1$\pm$0.5 &           3.0$\pm$0.4 &           2.0$\pm$0.4 &  \textbf{1.1$\pm$0.3} \\
\contacttag-t &    3.7$\pm$1.1 &    3.0$\pm$1.5 &           3.3$\pm$1.2 &           3.6$\pm$1.1 &  \textbf{1.4$\pm$0.5} \\
\ndctag-t     &    3.3$\pm$0.6 &    3.3$\pm$0.6 &           2.8$\pm$0.3 &           2.8$\pm$0.3 &  \textbf{2.7$\pm$0.6} \\
\tagtag-t     &    4.3$\pm$0.5 &    4.5$\pm$0.9 &           2.8$\pm$0.7 &           2.1$\pm$0.5 &  \textbf{1.2$\pm$0.5} \\
\threadtag-t  &    4.2$\pm$0.7 &    4.7$\pm$0.4 &           2.2$\pm$0.5 &           2.1$\pm$0.5 &  \textbf{1.8$\pm$0.6} \\
\dblptag-t    &    3.2$\pm$0.6 &    3.1$\pm$0.3 &           3.0$\pm$0.0 &           2.9$\pm$0.3 &  \textbf{2.8$\pm$0.6} \\
\bottomrule
\end{tabular}
\end{table}
A similar trend could be seen from the  rank-performance table of the \textit{micro} mode (Table~\ref{tab:micro_feat_rank_perf_auc}), where at least one combination involving \texttt{H} ranks higher than the rest in each row.
As compared with the analysis in the \textit{standalone} mode, where individual features were used, the \textit{micro} mode gives better scores; more so, when hypergraph features are involved.

\subsection{Macro Feature Combination Performances}
Finally, partitioning the features as per the \textit{macro} mode in Section~\ref{sec:exp:lp} gives us a total of five feature combinations, all of whose performances have been listed in Table~\ref{tab:macro_feat_perf_auc}.
The hypergraph based features perform much better with these feature combinations.
Even though \texttt{mac-H} underperforms the last two columns, compared with the purely graph oriented feature combinations (\textit{mac-G} and \textit{mac-W}), except for B-s, B-t, and D-t, it performs better.
\begin{table}
    \caption{XGBoost classification AUC scores for link prediction performed using various feature combinations: \texttt{G}, \texttt{W}, \texttt{H}, \texttt{GH}, \texttt{WH}}
    \label{tab:macro_feat_perf_auc}
    \centering
    \footnotesize
    \begin{tabular}{l@{\hspace{2mm}}c@{\hspace{2mm}}c@{\hspace{2mm}}c@{\hspace{2mm}}c@{\hspace{2mm}}c}
% \begin{tabular}{lccccc}
\toprule
{} &  \texttt{mac-G} &  \texttt{mac-W} &  \texttt{mac-H} &          \texttt{mac-GH} &          \texttt{mac-WH} \\
\midrule
\emailtag-s    &  93.10$\pm$1.00 &  93.06$\pm$1.05 &  93.89$\pm$0.41 &           93.90$\pm$0.52 &  \textbf{94.10$\pm$0.54} \\
\contacttag-s &  93.40$\pm$0.46 &  93.54$\pm$0.41 &  93.46$\pm$0.65 &           93.59$\pm$0.72 &  \textbf{93.68$\pm$0.44} \\
\ndctag-s     &  98.77$\pm$0.12 &  98.73$\pm$0.12 &  98.87$\pm$0.11 &           98.88$\pm$0.10 &  \textbf{98.89$\pm$0.15} \\
\tagtag-s     &  95.16$\pm$0.08 &  95.35$\pm$0.12 &  96.56$\pm$0.12 &  \textbf{96.60$\pm$0.09} &           96.56$\pm$0.11 \\
\threadtag-s  &  96.90$\pm$0.15 &  96.86$\pm$0.14 &  97.19$\pm$0.14 &  \textbf{97.20$\pm$0.16} &           97.19$\pm$0.15 \\
\dblptag-s    &  97.79$\pm$0.02 &  97.79$\pm$0.02 &  99.51$\pm$0.00 &  \textbf{99.52$\pm$0.00} &           99.51$\pm$0.00 \\
\midrule
\emailtag-t    &  74.44$\pm$1.50 &  78.29$\pm$1.64 &  79.05$\pm$2.14 &           79.56$\pm$1.83 &  \textbf{84.76$\pm$1.40} \\
\contacttag-t &  86.64$\pm$1.87 &  87.92$\pm$1.67 &  87.31$\pm$1.93 &           86.96$\pm$1.81 &  \textbf{88.46$\pm$1.68} \\
\ndctag-t     &  58.89$\pm$0.06 &  59.15$\pm$0.05 &  61.01$\pm$0.07 &           61.08$\pm$0.06 &  \textbf{61.41$\pm$0.07} \\
\tagtag-t     &  90.80$\pm$0.40 &  91.63$\pm$0.35 &  91.31$\pm$0.34 &           91.53$\pm$0.33 &  \textbf{92.23$\pm$0.30} \\
\threadtag-t  &  84.66$\pm$0.25 &  84.95$\pm$0.27 &  90.59$\pm$0.14 &           90.59$\pm$0.13 &  \textbf{90.77$\pm$0.15} \\
\dblptag-t    &  85.29$\pm$0.04 &  86.00$\pm$0.04 &  87.93$\pm$0.04 &           88.00$\pm$0.04 &  \textbf{88.44$\pm$0.05} \\
\bottomrule
\end{tabular}
\end{table}

\begin{table}
    \caption{AUC scores (\%) for structural (-s) and temporal (-t) link prediction using standalone features for \textit{contact-high-school} (\textit{i.e.}, dataset \contacttag). Row ids AA--Prn represent base predictors.}
    \label{tab:standalone_perf_auc_contact-high-school}
    \centering
    \footnotesize
    \begin{tabular}{l@{\hspace{2mm}}c@{\hspace{2mm}}c@{\hspace{2mm}}c@{\hspace{2mm}}c@{\hspace{2mm}}c@{\hspace{2mm}}c}
% \begin{tabular}{lcccccc}
\toprule
{} &         \texttt{std-G} &         \texttt{std-W} & \texttt{std-\Hmax} & \texttt{std-\Havg} & \texttt{std-\HLone} & \texttt{std-\HLtwo} \\
\midrule
AA-s  &  \textbf{93.0$\pm$0.5} &           92.8$\pm$0.3 &                      89.1$\pm$0.3 &                      92.1$\pm$0.3 &                     92.4$\pm$0.4 &                     92.6$\pm$0.4 \\
AS-s  &           91.2$\pm$0.3 &           88.1$\pm$0.2 &                      69.7$\pm$0.3 &                      91.9$\pm$0.3 &                     92.6$\pm$0.4 &            \textbf{92.8$\pm$0.4} \\
CN-s  &  \textbf{92.8$\pm$0.5} &           92.4$\pm$0.3 &                      77.3$\pm$0.4 &                      92.0$\pm$0.3 &                     92.2$\pm$0.4 &                     92.2$\pm$0.4 \\
Cos-s &  \textbf{92.8$\pm$0.4} &           \textbf{92.8$\pm$0.2} &                      77.9$\pm$0.3 &                      92.0$\pm$0.3 &                     92.4$\pm$0.4 &                     92.6$\pm$0.4 \\
PA-s  &  \textbf{63.6$\pm$0.6} &           62.0$\pm$0.8 &                      55.0$\pm$0.4 &                      56.0$\pm$1.2 &                     62.6$\pm$0.8 &                     62.1$\pm$0.9 \\
JC-s  &  \textbf{92.8$\pm$0.4} &           \textbf{92.8$\pm$0.3} &                      77.9$\pm$0.3 &                      92.0$\pm$0.3 &                     92.5$\pm$0.4 &                     92.7$\pm$0.4 \\
MxO-s &           92.6$\pm$0.4 &           92.6$\pm$0.3 &                      77.7$\pm$0.3 &                      92.0$\pm$0.3 &                     92.5$\pm$0.4 &            \textbf{92.7$\pm$0.4} \\
MnO-s &           \textbf{92.4$\pm$0.3} &           91.2$\pm$0.1 &                      77.3$\pm$0.4 &                      92.0$\pm$0.3 &                     92.3$\pm$0.4 &            \textbf{92.4$\pm$0.4} \\
NM-s  &  \textbf{92.8$\pm$0.4} &           92.7$\pm$0.3 &                      77.9$\pm$0.3 &                      92.0$\pm$0.3 &                     92.5$\pm$0.4 &                     92.6$\pm$0.4 \\
Prn-s &           90.6$\pm$0.4 &           90.1$\pm$0.2 &                      77.9$\pm$0.3 &                      92.0$\pm$0.3 &                     92.4$\pm$0.4 &            \textbf{92.6$\pm$0.4} \\
\midrule
AA-t  &           87.9$\pm$0.3 &           87.8$\pm$0.3 &                      83.5$\pm$0.2 &             \textbf{88.0$\pm$0.2} &                     86.7$\pm$0.3 &                     87.0$\pm$0.3 \\
AS-t  &           87.5$\pm$0.2 &           84.8$\pm$0.3 &                      67.1$\pm$0.3 &             \textbf{88.0$\pm$0.2} &                     87.1$\pm$0.3 &                     87.5$\pm$0.3 \\
CN-t  &           87.7$\pm$0.3 &           87.4$\pm$0.3 &                      72.2$\pm$0.3 &             \textbf{87.9$\pm$0.2} &                     86.5$\pm$0.3 &                     86.5$\pm$0.3 \\
Cos-t &  \textbf{88.5$\pm$0.2} &  \textbf{88.5$\pm$0.2} &                      73.3$\pm$0.3 &                      88.0$\pm$0.2 &                     86.8$\pm$0.3 &                     87.1$\pm$0.3 \\
PA-t  &  \textbf{53.8$\pm$0.3} &  \textbf{53.8$\pm$0.3} &                      51.2$\pm$0.4 &                      50.3$\pm$0.4 &                     52.5$\pm$0.4 &                     52.3$\pm$0.4 \\
JC-t  &  \textbf{88.4$\pm$0.2} &  \textbf{88.4$\pm$0.2} &                      73.2$\pm$0.3 &                      88.0$\pm$0.2 &                     86.9$\pm$0.3 &                     87.2$\pm$0.3 \\
MxO-t &           87.9$\pm$0.2 &  \textbf{88.0$\pm$0.2} &                      72.6$\pm$0.3 &             \textbf{88.0$\pm$0.2} &                     86.9$\pm$0.3 &                     87.2$\pm$0.3 \\
MnO-t &  \textbf{88.2$\pm$0.2} &           87.1$\pm$0.2 &                      72.2$\pm$0.3 &                      88.0$\pm$0.2 &                     86.7$\pm$0.3 &                     86.8$\pm$0.3 \\
NM-t  &  \textbf{88.3$\pm$0.2} &           88.2$\pm$0.2 &                      73.2$\pm$0.3 &                      88.0$\pm$0.2 &                     86.8$\pm$0.3 &                     87.1$\pm$0.3 \\
Prn-t &           85.8$\pm$0.2 &           85.0$\pm$0.2 &                      73.3$\pm$0.3 &             \textbf{88.0$\pm$0.2} &                     86.8$\pm$0.3 &                     87.1$\pm$0.3 \\
\bottomrule
\end{tabular}
\end{table}

\begin{table}
    \caption{AUC scores (\%) for structural (-s) and temporal (-t) link prediction using micro-feature-combination via XGBoost for \textit{contact-high-school} (\textit{i.e.}, dataset \contacttag). Row ids AA--Prn represent base predictors.}
    \label{tab:micro_feat_perf_auc_contact-high-school}
    \centering
    \footnotesize
    
\end{table}
\vspace{-0.5cm}
\subsection{Standalone Feature Performances}
For link prediction experiments in the \textit{standalone} mode (Section~\ref{sec:exp:lp}), we show results only for a single dataset: \textit{contact-high-school} (dataset~\contacttag) in Table~\ref{tab:standalone_perf_auc_contact-high-school}.

Although we did not expect to do better than the baselines in the standalone mode, since individual hypergraph scores might not be powerful link predictors, yet we observe decent performances in the last four columns (the only ones that correspond to hypergraph-based scores).
Going by a base predictor individually (row-wise), graph-versions (\texttt{std-G}) of Adamic Adar (AA) for structural- and Cosine Similarity (Cos) for the temporal-mode perform best.

We consolidate these results for all datasets by finding the mean (over all base predictors) ``rank'' among all standalone modes (\texttt{std-G}, \texttt{std-W}, \texttt{std-\Hmax}, \texttt{std-\Havg}, \texttt{std-\HLone}, \texttt{std-\HLtwo}) in Table~\ref{tab:standalone_rank_perf_auc}.
This is how it has to be interpreted: for example, for dataset A, in the structural mode rank of \texttt{std-G} being $1.3\pm 0.6$ means out of the six standalone modes, \texttt{std-G} stands at a mean position of $1.3$ (with variance 0.6), when evaluated across all base predictors.

\begin{table}
    \caption{Rank-performances \textit{w.r.t.} AUC scores from Table~\ref{tab:standalone_perf_auc_contact-high-school} across all datasets. Row ids A--F represent dataset ids (ref. Section~\ref{sec:exp:datasets}), where -s and -t refer to structural and temporal respectively.}
    \label{tab:standalone_rank_perf_auc}
    \centering
    \footnotesize
    \begin{tabular}{l@{\hspace{2.5mm}}c@{\hspace{2.5mm}}c@{\hspace{2.5mm}}c@{\hspace{2.5mm}}c@{\hspace{2.5mm}}c@{\hspace{2.5mm}}c}
\toprule
{} &        \texttt{std-G} &        \texttt{std-W} & \texttt{std-\Hmax} & \texttt{std-\Havg} & \texttt{std-\HLone} & \texttt{std-\HLtwo} \\
\midrule
\emailtag-s    &  \textbf{1.3$\pm$0.6} &           3.8$\pm$1.0 &                       3.0$\pm$1.5 &                       3.2$\pm$1.3 &                      5.4$\pm$1.3 &                      4.3$\pm$1.0 \\
\contacttag-s &  \textbf{1.8$\pm$1.2} &           3.2$\pm$1.3 &                       6.0$\pm$0.0 &                       4.5$\pm$0.8 &                      3.2$\pm$0.9 &                      2.2$\pm$1.0 \\
\ndctag-s     &           3.2$\pm$1.1 &           4.1$\pm$1.3 &              \textbf{1.4$\pm$1.2} &                       5.8$\pm$0.6 &                      3.7$\pm$1.0 &                      2.8$\pm$0.9 \\

\tagtag-s     &           4.0$\pm$1.1 &           4.3$\pm$1.6 &                       3.4$\pm$0.8 &                       5.7$\pm$0.6 &                      2.2$\pm$0.6 &             \textbf{1.4$\pm$0.7} \\
\threadtag-s  &           4.1$\pm$1.3 &           3.8$\pm$0.9 &                       3.0$\pm$1.1 &                       5.8$\pm$0.6 &                      \textbf{2.2$\pm$1.0} &             \textbf{2.2$\pm$0.7} \\
\dblptag-s    &           3.4$\pm$0.4 &           3.4$\pm$0.2 &                       3.6$\pm$0.4 &                       3.8$\pm$0.8 &             \textbf{3.2$\pm$0.8} &                      3.6$\pm$0.2 \\
\midrule
\emailtag-t    &           2.9$\pm$0.7 &  \textbf{1.6$\pm$1.2} &                       3.8$\pm$1.2 &                       2.1$\pm$0.7 &                      5.8$\pm$0.4 &                      4.8$\pm$0.4 \\
\contacttag-t &  \textbf{2.0$\pm$0.9} &           2.7$\pm$1.3 &                       5.9$\pm$0.3 &                       2.2$\pm$1.5 &                      4.4$\pm$0.8 &                      3.6$\pm$0.8 \\
\ndctag-t     &           3.2$\pm$1.4 &           3.1$\pm$1.6 &              \textbf{2.6$\pm$1.0} &                       4.4$\pm$0.9 &                      4.1$\pm$1.2 &                      3.5$\pm$1.0 \\
\tagtag-t     &           4.0$\pm$0.9 &           4.0$\pm$1.6 &                       3.6$\pm$0.9 &                       5.8$\pm$0.6 &             \textbf{1.8$\pm$0.5} &                      \textbf{1.8$\pm$0.7} \\
\threadtag-t  &           4.1$\pm$1.3 &           3.6$\pm$1.1 &                       3.6$\pm$0.9 &                       5.8$\pm$0.6 &             \textbf{1.6$\pm$0.8} &                      2.2$\pm$0.6 \\
\dblptag-t    &           3.4$\pm$0.4 &  \textbf{3.2$\pm$0.8} &                       3.6$\pm$0.4 &                       3.8$\pm$0.8 &                      3.4$\pm$0.2 &                      3.6$\pm$0.2 \\
\bottomrule
\end{tabular}
\end{table}

% As described in Section~\ref{sec:exp:lp}, we use the newly constructed scores to predict links between unlinked pairs.
% For completeness' sake, we deal with both structural and temporal version of the problem, and report our results on datasets described in Section~\ref{sec:exp:datasets}.

% \subsubsection{Structural}
% \label{sec:results:lp:structural}
% \input{sections/results/link_prediction/structural.tex}

% \subsubsection{Temporal}
% \label{sec:results:lp:temporal}
% \input{sections/results/link_prediction/temporal.tex}

 \section{Conclusion and Future Work}
  \label{sec:conc}
  Structural (topological) node similarity scores have a long history in similarity computation, and have been equally successful as well.
Also, hypergraph networks are very frequently used in works involving similarity computation, albeit not being exploited for the task per se.
We set out to use the underlying hypergraph structure of networks to generate new features for similarity computation.
Apart from establishing a strong theoretical foundation by devising functional templates that could help standard similarity computation scores getting translated from graphs to hypergraphs, we are also able to elucidate hypergraphs' contribution in predicting links.
We perform a number of experiments to show the importance of using hypergraph-based topological features for similarity computation, including showing a mutual-information based perspective.
A few take-away messages are:
\begin{enumerate}
    \item Higher-order structure does have richer information than graphs.
    \item When available, using the underlying hypergraph structure would term fruitful in link prediction.
    \item Various matrix norms combine hyperedge information in different ways; the best bet is to use multiple norms and choose the best.
    \item Unless the similarity computation model overfits, all hypergraph features should be used, if possible.
\end{enumerate}

As a next step, we would like to use the functional-formulation for global, random-walk based measures.

\bibliographystyle{plain}
\bibliography{ms}
\end{document}